\documentclass[aps,prl,twocolumn,showpacs,superscriptaddress,groupedaddress]{revtex4}  

\usepackage{amssymb}
\usepackage{amsmath}
\usepackage{epsfig}
\usepackage{epstopdf}
\usepackage{bm}
\usepackage{graphicx,epsfig}
\usepackage{mathrsfs}
\usepackage{dcolumn}
\usepackage{color}
\usepackage{natbib}
\usepackage{CJK}
\usepackage{soul}
\usepackage{miniplot}
\def\be{\begin{equation}}
\def\ee{\end{equation}}
\def\bea{\begin{eqnarray}}
\def\eea{\end{eqnarray}}

\def\ra{\rangle}

\allowdisplaybreaks[2]

\begin{document}

\title{High-frequency Light Reflector via Low-frequency Light Control}
\author{Da-Wei Wang}
\affiliation{Texas A$\&$M University, College Station, TX 77843, USA}
\author{Shi-Yao Zhu}
\affiliation{Beijing Computational Science Research Centre, Beijing 100084, China}
\author{J\"org Evers}
\affiliation{Max-Planck-Institut f\"ur Kernphysik, Saupfercheckweg 1, 69117 Heidelberg, Germany}
\author{Marlan O. Scully}
\affiliation{Texas A$\&$M University, College Station, TX 77843, USA}
\affiliation{Princeton University, Princeton, New Jersey 08544, USA}
\affiliation{Baylor University, Waco, TX 76706, USA}

\date{\today }

\begin{abstract}
We show that the momentum of light can be reversed via the atomic coherence created by another light with one or two orders of magnitude lower frequency. Both the backward retrieval of single photons from a timed Dicke state and the reflection of continuous waves by high-order photonic band gaps are analysed. The required control field strength scales linearly with the nonlinearity order, which is explained by the dynamics of superradiance lattices. Experiments are proposed with $^{85}$Rb atoms and Be$^{2+}$ ions. This holds promise for light-controllable X-ray reflectors.
\end{abstract}


\maketitle

\emph{Introduction}---Photonic crystals (PCs)~\cite{Yablonovitch1987,John1987} can control the flow of light with the well-known phenomenon of photonic band gaps (PBGs). The fabrication of PCs requires high accuracy on the wavelength scale, which renders the formation of band gaps at higher photon frequencies and shorter wavelengths difficult. Techniques like focused-ion-beam etching and other micro-fabrication techniques already make ultraviolet PCs accessible \cite{Wu2004, Ganesh2006, Radeonychev2009}. But advancing to the extreme ultraviolet (XUV) or X-ray regime remains challenging. Nevertheless, novel methods to control the flow of high-frequency light are still desirable, to complement recent progress in X-ray optics such as the development of X-ray mirrors based on diamonds \cite{Shvydko2011}.

Besides physical PCs, electromagnetically induced transparency (EIT) \cite{Boller1991} has been used to form optically controllable PBGs~\cite{Andre2002,Artoni2006}. But in the commonly used $\Lambda$-type rubidium and cesium three-level systems, the driving and the probe light fields have near-degenerate wavelengths and therefore only the first order band gap exists. Since strong control fields are lacking at X-ray photon energies, it is experimentally favorable to control the weak probe light of short wavelength with a strong control field of long wavelength. Although the control of the transmission of XUV \cite{Ranitovic2011} or even X-ray \cite{Glover2010} light based on intense optical control laser fields has been realized, their reflection which involves high-order photonic band gaps (HOPBG) \cite{Straub2003, Barillaro2007, Radeonychev2009, Morozov2010, Lu2012} remains unsolved.

\begin{figure}[t]
    \epsfig{figure=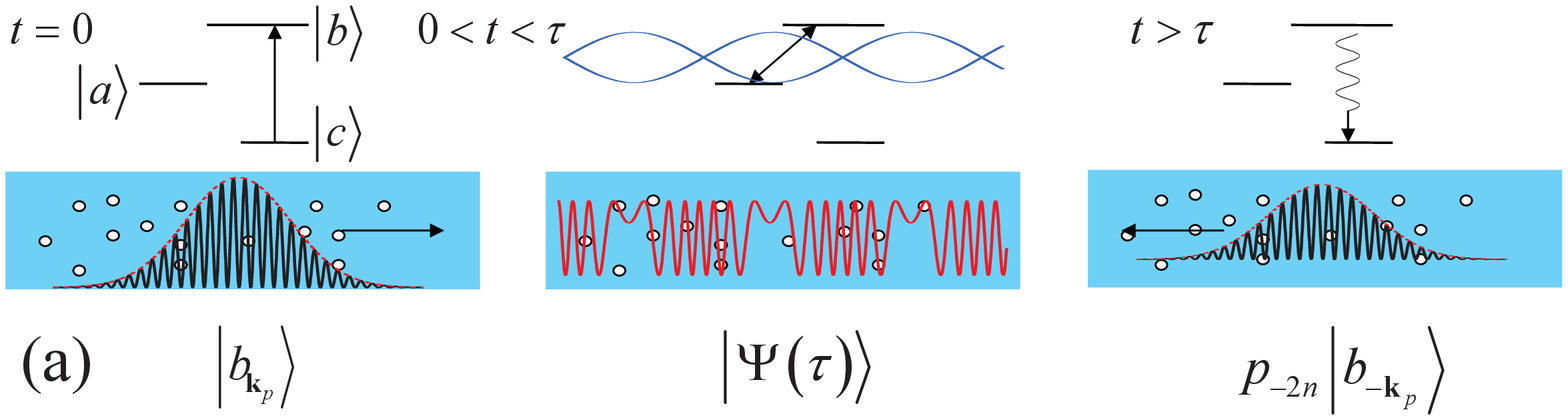, angle=0, width=0.48\textwidth}
    \epsfig{figure=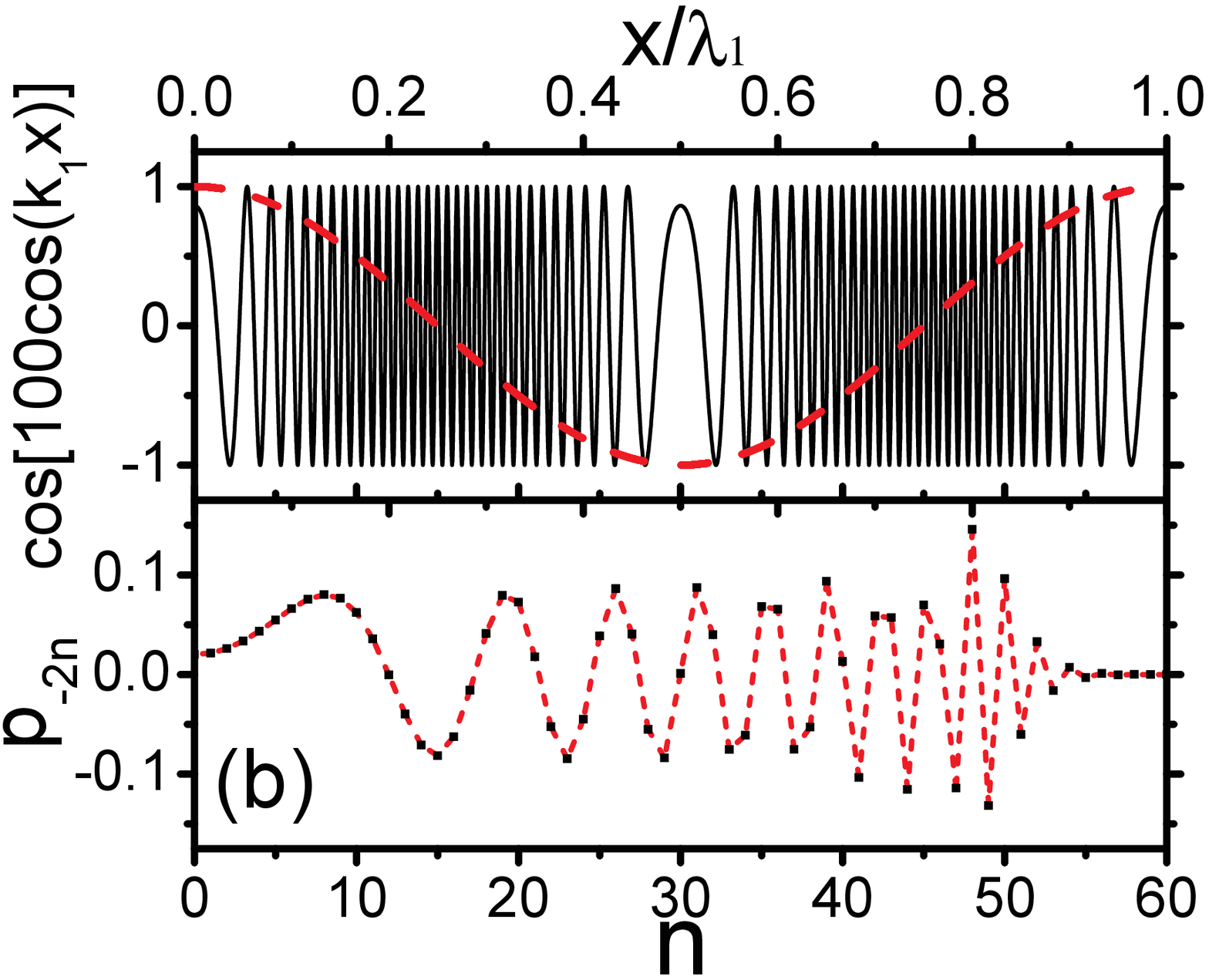, angle=0, width=0.23\textwidth}
    \epsfig{figure=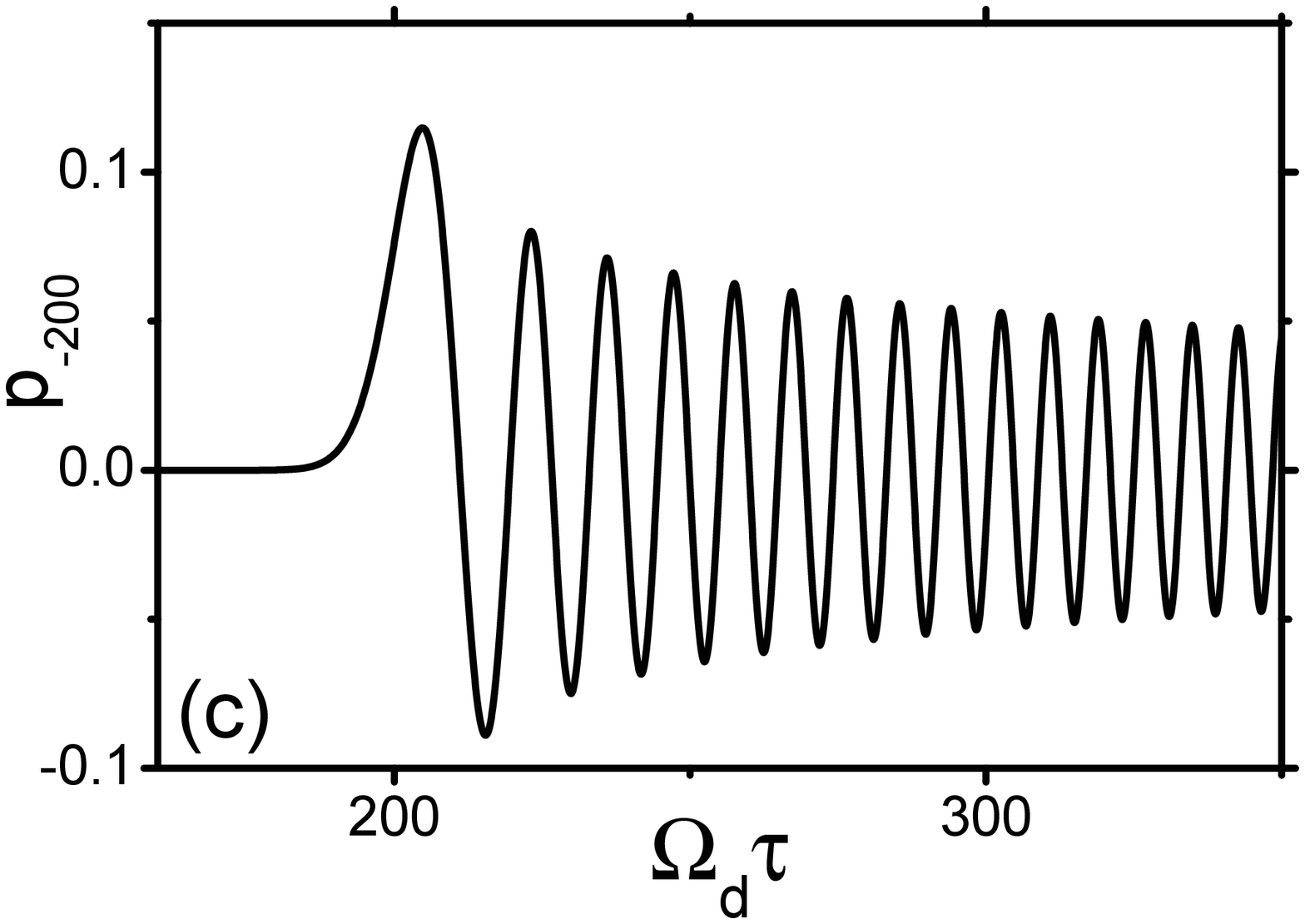, angle=0, width=0.23\textwidth}
\caption{(Color online) $(a)$ Coherent modulation of the timed Dicke state. Left: The atomic ensemble is collectively excited to the timed Dicke state $\left|b_{\mathbf{k}_p}\right\rangle$ at $t=0$ by a pulse with short wavelength $\lambda_p=\lambda_{bc}$ and wave vector $\mathbf{k}_p$. Middle: A standing wave with long wavelength $\lambda_1=\lambda_{ba}=n\lambda_{bc}$ induces Rabi cycles between $\left|b\right\rangle$ and $\left|a\right\rangle$. 
The wave pattern shows the spatial population modulation of $\left|b\right\rangle$, which exhibits much finer structures than the standing wave. Right: After time $\tau$, $\left|b_{\mathbf{k}_p}\right\rangle$ is transferred to the reverse timed Dicke state $\left|b_{\mathbf{-k}_p}\right\rangle$ with probability $p_{-2n}^2$ and collectively emits a photon in the backward direction $\mathbf{-k}_p$. $(b)$ Upper: The spatial population modulation (black solid) of $\left|b\right\rangle$ by the standing wave pattern (red dash). Lower: The probability amplitude $p_{-2n}$. $\Omega_d\tau=100$. (c) The probability amplitude $p_{-200}$ as a function of $\Omega_d\tau$.}
\label{scheme}
\end{figure}

In this Letter, we show the reflection of high frequency photons by the high order nonlinearity of a low-frequency standing wave coupled EIT scheme. The key feature of the high-order nonlinearity involved is that the required field strength scales linearly with the nonlinearity order, in contrast to the power law dependence in common nonresonant nonlinear media. We consider the $\Lambda$-type EIT scheme, as shown in Fig.~\ref{scheme} (a). A probe field couples the ground state $\left|c\right\rangle$ to the excited state $\left|b\right\rangle$. A standing wave control field couples $\left|b\right\rangle$ to a meta-stable state $\left|a\right\rangle$. The Rabi frequency of the forward (backward) component of the standing wave is ${{\Omega }_{1}}$ (${{\Omega }_{2}}$). If the wavelength of the coupling field is $n$ times the one of the probe field, and the decoherence time of a probe photon excitation is ${{\tau }_{bc}}$, the requirement of effective reflection of the probe field is
\begin{equation}
{{\tau }_{bc}}{{\Omega }_{1,2}}>n.
\label{criteria}
\end{equation}
This relation can be understood from the momentum conservation. To reverse the momentum of the probe photon, the ensemble should emit $n$ coupling photon in the forward mode and absorb $n$ coupling photon from the backward mode. The time cost in one cycle of emission and absorption is $1/\Omega_{1,2}$. The total process should be completed within the decoherence time, $n/\Omega_{1,2}<\tau_{bc}$, which leads to Eq. (\ref{criteria}). This linear dependence between the order of the nonlinearity and the field strength will be confirmed in the following for both single photons and continuous wave probe fields. The hidden physics of this nonlinearity is envisioned and explained with a momentum space Fano lattice \cite{Miroshnichenko2010}, the superradiance lattice \cite{Wang2015}.

\emph{Single photons}---We first apply the idea to the backward retrieval of a single photon from a medium prepared at time $t=0$ in the timed Dicke state of an ensemble of $N$ atoms by absorption of a single probe photon,
\begin{equation}
\left|b_{\mathbf{k}_p}\right>=\frac{1}{\sqrt{N}}\sum\limits_{j=1}^{N}\exp\left(ik_p x_j\right)\left|c_1,c_2,...,b_j,...c_N\right>,
\label{Psi}
\end{equation}
where $\mathbf{k}_p$ is along the $+\hat{x}$ direction. For an ensemble large compared to the probe wavelength $\lambda_p$, this state will decay with the collective decoherence time $\tau_{bc}$ and emit a photon in $+\hat{x}$ direction \cite{Scully2006}. In order to retrieve a photon in $-\hat{x}$ direction, we apply a coherent standing wave to drive the transition between $\left| b \right\rangle$ and $\left|a\right\rangle $ in the time scale $\tau<\tau_{bc}$ directly after the excitation. The wave vector of the forward (backward) component of the standing wave is $\mathbf{k}_1=k_1 \hat{x}$ ($\mathbf{k}_2=-\mathbf{k}_1$). With the interaction Hamiltonian $H=-\sum_{j=1}^{N}\hbar\Omega_d \text{cos}(k_1x_j)\sigma_j^x$ where $\Omega_d=2\Omega_1=2\Omega_2$ and $\sigma_j^x=\left|b_j\right\rangle\left\langle a_j\right|+h.c.$, the evolution operator $U\left(t\right)=\exp\left(-iHt/\hbar\right)$ is then
$
U\left(t\right)=\sum_{j=1}^{N}\cos\left[ {{\Omega }_{d}}t\text{cos}({{k}_{1}}{{x}_{j}}) \right]I+i\sin\left[ {{\Omega }_{d}}t\text{cos}({{k}_{1}}{{x}_{j}}) \right]\sigma_j^x
$
where $I$ is the $2\times 2$ unit matrix. After time $\tau$, the wave function is
\begin{eqnarray}
\begin{aligned}
\left|\Psi\left( \tau \right) \right\rangle=U\left(\tau\right)\left|b_{\mathbf{k}_p}\right\rangle.
\end{aligned}
\label{psit}
\end{eqnarray}
The projection of $\left|\Psi\left(\tau\right)\right\rangle$ on the target state associated to the $n$th-order nonlinear process $\left|b_{\mathbf{k}_p+2n\mathbf{k}_1}\right\rangle=\frac{1}{\sqrt{N}}\sum_j\exp\left[i(k_p+2nk_1)x_j\right]\left|c_1,c_2,...,b_j,...c_N\right>$ is
\begin{equation}
p_{2n}=\left\langle b_{\mathbf{k}_p+2n\mathbf{k}_1}|\Psi\left(\tau\right)\right\rangle=(-1)^n J_{2n}(\Omega_d\tau),
\label{p2n}
\end{equation}
where $J_n(x)$ is the Bessel function. Eq. (\ref{p2n}) is a reminiscence of one-dimensional tight-binding lattices, which will be discussed latter. 

In the upper part of Fig.~\ref{scheme} (b), we plot the population modulation in Eq.~(\ref{psit}), which has much finer spatial structure than the standing wave. This fine structure has been used in sub-wavelength lithography \cite{Liao2010} and lies at the heart of our analysis. After time $\tau$, the atomic ensemble has a probability $p_{-2n}^2$ to be in state $\left|b_{\mathbf{k}_p-2n\mathbf{k}_1}\right\rangle$. If $\mathbf{k}_p=n\mathbf{k}_1$, the reversely timed Dicke state $\left|b_{-\mathbf{k}_p}\right\rangle$ is obtained, and the single photon is superradiantly emitted in the backward direction. The lower part of Fig.~\ref{scheme} (b) shows the related cut off order $n_c\approx\frac{1}{2}\Omega_d\tau=\Omega_1\tau < \Omega_1\tau_{bc}$, which proves the criteria in Eq.~(\ref{criteria}). As an example for $\mathbf{k}_p=100\mathbf{k}_1$, the probability amplitude $p_{-200}$ of $\left|b_{-\mathbf{k}_p}\right\rangle$ is shown in Fig.~\ref{scheme} (c) as a function of $\Omega_d\tau$. The maximum probability of the backward retrieval of the photon reaches to 1\% around $\Omega_d\tau=205$. 

The requirement in Eq. (\ref{criteria}) sets a stringent restriction if $\tau_{bc}$ is small especially for XUV and X-ray. Fortunately, the above mechanism can be extended to a decoherence-free configuration with Raman transitions \cite{Wang2014a}. Furthermore, continuous driving fields can be replaced by $\pi$-pulses, which allow us disentangling various quantum paths to transport the initial state to the target state.

\emph{Continuous wave}---In the following, we will investigate the backward reflection for continuous probe waves. The dynamics can be explicitly calculated from the EIT susceptibility \cite{Artoni2006}
\begin{equation}
          \begin{aligned}
           \chi\left( x \right)
           &=-3\pi \mathcal{N}\frac{{{\Gamma }_{bc}}}{{{\gamma
           }_{bc}}}\frac{{{\gamma }_{bc}}}{{{-\Delta }_{p}}+i{{\gamma }_{bc}}
           -\frac{{{\left| {{\Omega }_{1}}{{e}^{i{{k}_{1}}x}}+{{\Omega
           }_{2}}{{e}^{-i{{k}_{1}}x}} \right|}^{2}}}{{{\Delta }_{2\text{ph}}}+i{{\gamma
           }_{ac}}}}\\
           &=\sum\limits_{m}{{{\chi }^{\left( 2m+1
           \right)}}{{e}^{-2mi{{k}_{1}}x}}},
          \end{aligned}
\end{equation}
where $\mathcal{N}$ is the number of atoms in the volume $\left(\lambda_{bc}/2\pi\right)^3$  and $\Gamma_{bc}$ is the radiative decay rate from $\left| b \right\rangle$ to $\left| c \right\rangle $. $\gamma _{ij}$ and $\omega_{ij}$ are the dephasing rate and the transition frequency between $\left| i \right\rangle$ and $\left|j\right\rangle$. The two driving light fields have the same frequency $\nu_{1}$ but opposite wave vectors  $\mathbf{k}_1=-\mathbf{k}_2=\hat{x}{\nu_1}/c$. $\Delta_{1}=\omega_{ba}-\nu_1$ is the detuning of the driving field. $\Delta_{p}={\omega }_{bc}-\nu_{p}$ and $\Delta_{2\text{ph}}=\Delta_1-\Delta_p=\omega_{ca}+\nu _{p}-\nu_{1}$ are the one- and two-photon detunings of the probe field. $\chi^{\left(2m+1\right)}$ is the Fourier component of $\chi \left(x\right)$ with phase ${e}^{-2mi{{k}_{1}}x}$. 

\begin{figure}[t]
\epsfig{figure=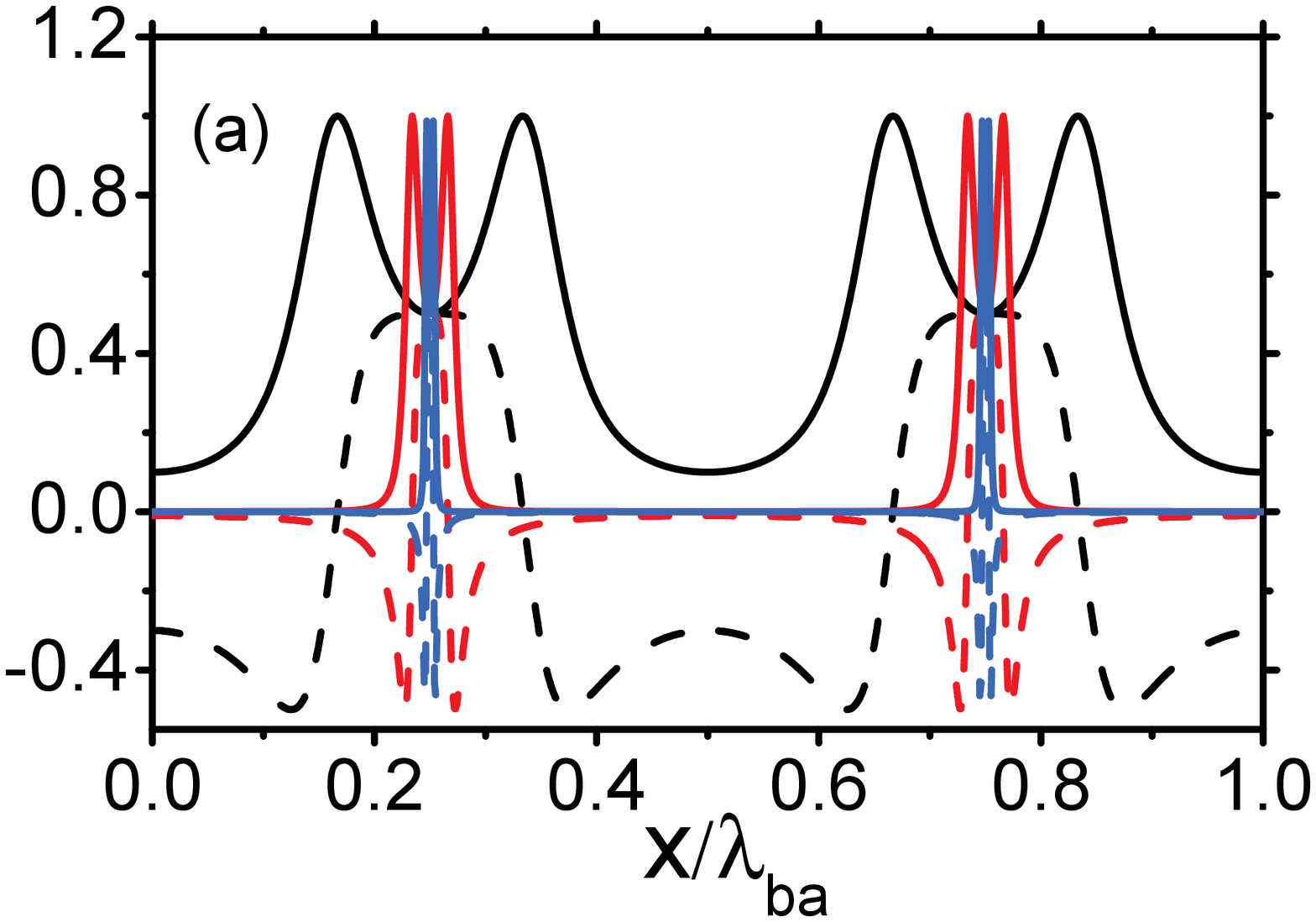, angle=0, width=0.23\textwidth}
\epsfig{figure=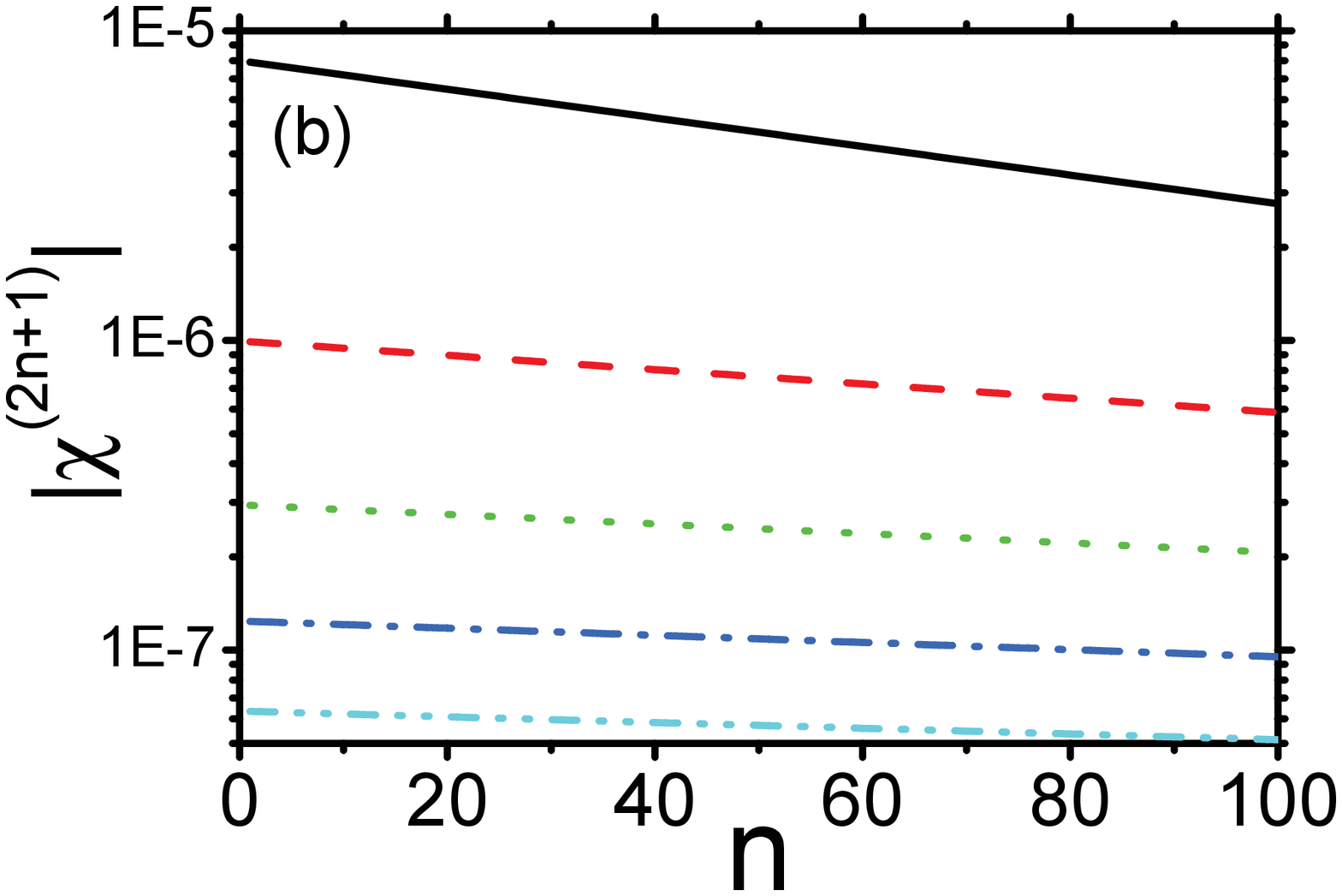, angle=0, width=0.23\textwidth}
\caption{(Color online) (a) The imaginary (solid line) and the real (dash line) parts of the spatially periodic susceptibility are plotted in units of $3\pi \mathcal{N}\Gamma_{bc}/\gamma_{bc}$. $\Delta_p=\gamma_{bc}$. $\gamma_{ac}=0$, ${{\Omega }_{1}}={{\Omega }_{2}}={{\gamma }_{bc}}$ (black),
$5{{\gamma }_{bc}}$ (red) and $25{{\gamma }_{bc}}$ (blue). (b) The magnitude of the $n$th-order Fourier components of the susceptibility ${{\chi }^{\left( 2n+1 \right)}}$ in units of $3\pi \mathcal{N}\Gamma_{bc}/\gamma_{bc}$. ${{\gamma }_{ac}}=0$, $\Delta_1=0$, ${{\Delta }_{p}}=0.1{{\gamma }_{bc}}$. ${{\Omega }_{1}}={{\Omega }_{2}}=20{{\gamma }_{bc}}$ (solid), $40{{\gamma }_{bc}}$ (dash), $60{{\gamma }_{bc}}$ (dot), $80{{\gamma }_{bc}}$ (dash dot) and $100{{\gamma }_{bc}}$ (dash dot dot).} 
\label{FC}
\end{figure}

In Fig.~\ref{FC} (a), we plot the real and the imaginary parts of $\chi \left( x \right)$. The susceptibility is periodically modulated. One interesting feature is that the modulation is sharply concentrated at the nodes of the standing wave as ${{\Omega }_{1,2}}$ is much larger than ${{\gamma }_{bc}}\equiv {1}/{{{\tau }_{bc}}}\;$, i.e., ${{\Omega }_{1,2}}{{\tau }_{bc}}\gg 1$. This is related to the requirement in Eq. (\ref{criteria}). The reason is that under the condition ${{\Omega }_{1,2}}{{\tau }_{bc}}\gg 1$, the periodic $\delta $-function like susceptibility has slowly decaying high-order components $\chi^{\left( 2m+1 \right)}$ which contribute to high-order Bragg reflection.

The Fourier components are calculated explicitly,
\begin{equation}
\chi^{\left( 2m+1 \right)}=A{{z}^{m}},
\end{equation}
where $z=\left[-B+\text{sign}\left( m \right)\sqrt{{{B}^{2}}-4}\right]/{2}$, $A=3\pi\mathcal{N} {{{\Gamma }_{bc}}\left( {{\Delta }_{2\text{ph}}}+i{{\gamma }_{ac}} \right)}/({{{\Omega }_{1}}{{\Omega }_{2}}}{\sqrt{{{B}^{2}}-4}})$ and $B=\left[\Omega _{1}^{2}+\Omega _{2}^{2}-\left( -{{\Delta }_{p}}+i{{\gamma }_{bc}} \right)\left( {{\Delta }_{2\text{ph}}}+i{{\gamma }_{ac}} \right) \right]/(\Omega_{1}\Omega_{2})$. If ${\Omega}_{1}={\Omega}_{2}$ exceeds all detunings and decoherence rates, $B\approx 2$ and $z\approx 1$. In this case, the absolute values of successive orders of the susceptibility, $\left| z \right|\approx 1$, are approximately the same, and high-order components significantly contribute to the susceptibility. In Fig.~\ref{FC} (b), we plot the magnitude of the Fourier components as a function of the order $n$ for different driving field strengths. For example, we find $|\chi^{(201)}/\chi^{(1)}|=0.35$ for $\Omega_1=20\gamma_{bc}$. In contrast, $|\chi^{(201)}/\chi^{(1)}|=0.81$ for $\Omega_1=100\gamma_{bc}$.

Near the phase matching condition ${{\mathbf{k}}_{p}}-2n{{\mathbf{k}}_{1}}=-{{\mathbf{k}}_{p}}$, a two-mode approximation is justified, and we consider the probe mode ${\mathbf{k}}_{p}$ and the $n$th order Bragg mode $-{\mathbf{k}}_{p}$ generated by the $(2n+1)$th order coherence $\chi^{\left(2n+1 \right)}$ only. Their slowly varying amplitudes ${\mathcal{E}}_{p}$ and ${\mathcal{E}}_{e}$ are governed by the following equations
\begin{equation}
\begin{aligned}
&\frac{\partial}{\partial x}{{\mathcal{E}}_{p}}=-\beta{{\mathcal{E}}_{p}}+i\kappa^{(2n+1)}{{\mathcal{E}}_{e}}{{e}^{-i\Delta {{k}_{n}}x}},\\
&\frac{\partial }{\partial x}{{\mathcal{E}}_{e}}=\beta {{\mathcal{E}}_{e}}-i\kappa^{(2n+1)}{{\mathcal{E}}_{p}}{{e}^{i\Delta {{k}_{n}}x}}.
\end{aligned}
\label{coupled equations}
\end{equation}
Here, ${k}_p=\nu_p\sqrt{1+\chi^{(1)}}/c$ is the magnitude of $\mathbf{k}_p$, $\theta$ is the angle between $\mathbf{k}_p$ and $\mathbf{k}_1$, $\Delta k_n=2k_p\cos\theta-2nk_1$ the wave vector mismatch, $\beta ={\nu}_{p}^2\operatorname{Im}{{\chi }^{\left( 1 \right)}}/2k_pc^2\cos \theta$ the depletion rate, and $\kappa^{(2n+1)}={\nu}_{p}^2{\chi }^{\left( 2n+1 \right)}/2k_pc^2\cos \theta$ is the coupling coefficient. The reflectance $R$ can be calculated analytically from these two equations~\cite{Wang2013}. For an infinitely long sample, we find \cite{Wang2014}
\begin{equation}
R={{\left| \frac{\sqrt{1-{{u}^{2}}}-1}{u} \right|}^{2}}
\end{equation}
where $u=\kappa^{(2n+1)}/(\Delta k_n/2+i\beta)$. $R$ increases the fastest with $\left|u\right|$ along $u$'s real axis and approaches 1 when $|u|\ge 1$ where band gaps appear. In other radial directions, $R$ also increases with $|u|$, and the slowest gradient is along the imaginary axis. Near the phase matching condition  $\Delta k_n\approx 0$, $u\approx \chi^{(2n+1)}/\chi^{(1)}=z^n$. For strong driving fields and near the EIT point, $\Omega_1=\Omega_2\gg \gamma_{bc}\gg\Delta_p$, we have $z\approx-1+2\sqrt{i\gamma_{bc}\Delta_p/\Omega _{1}^{2}}$. Then $u\approx (-1)^n \left(1- 2n\sqrt{i\gamma_{bc}\Delta_p/\Omega _{1}^{2}}\right)$ and the large reflectivity requires
\begin{equation}
\frac{\Omega _{1}^{2}}{\gamma_{bc}\Delta_p}\gg n^2.
\label{bgr}
\end{equation}
This confirms the intuitive requirement Eq.~(\ref{criteria}) for high-order coherence. Operation close to the EIT point $\Delta_p\ll \Omega_1/n$ is required to reduce the absorption induced by all other orders of the coherence.

\begin{figure}[t]
\begin{center}
\epsfig{figure=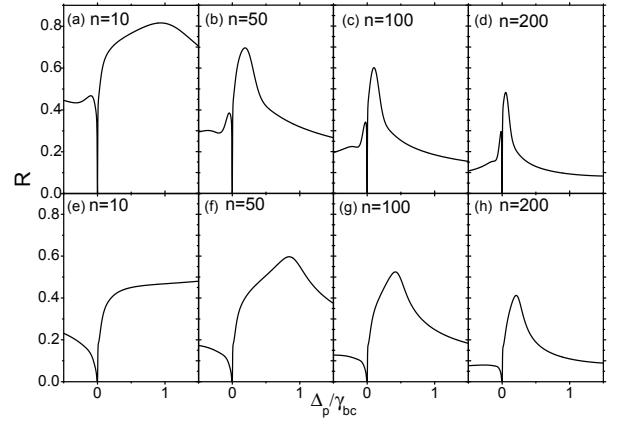, angle=0, width=0.45\textwidth}
\end{center}
\caption{The reflectivity due to $n$’th order Bragg reflection. ${{\Omega }_{1}}={{\Omega }_{2}}=200{{\gamma }_{bc}}$, ${{\gamma }_{ac}}=0$, $\Delta_{1}=0$. (a)-(d), $\Delta k_{n}^{0}={{10}^{-5}}{{{\nu }_{p}}}/{c}\;$; (e)-(h), $\Delta k_{n}^{0}=5\times {{10}^{-5}}{{{\nu }_{p}}}/{c}\;$. The sample length is ${{10}^{5}}{{\lambda }_{bc}}$. The density is $\mathcal{N}=0.01$. $\Gamma_{bc}=\gamma_{bc}$. The sample length is $10^{5}\lambda_{bc}$.}
\label{bg_cold}
\end{figure} 
In Fig.~\ref{bg_cold}, we plot the reflectivity $R$ for parameters approximately satisfying the $n$th-order Bragg condition for different $n$. The wavevector mismatches in free vacuum $\Delta k_{n}^{0}=2{\left( {{\nu }_{p}}\cos \theta -n{{\nu }_{1}} \right)}/{c}\;$ can be tuned by the incidence angle $\theta $. The dispersion contribution to the wave vector mismatch $\Delta k_n-\Delta k_{n}^{0}$ determines on which side of the transparency point the band gap appears. Here the band gap is characterized by a high reflection  plateau or peak \cite{Andre2002}. With increasing $n$, the band gaps shrink slowly and approaches the transparency point, which confirms the requirement Eq.~(\ref{bgr}). By changing $\theta$ and thus $\Delta k_{n}^{0}$, the position and the width of the band gap can be tuned via the compensation of the dispersion induced phase mismatch, as can be seen by comparing the two rows in Fig.~\ref{bg_cold}. 

\emph{Experimental Realization}---To evaluate the feasibility of an implementation, the robustness of the band gap against decoherence and inhomogeneous broadening (e.g., Doppler effect) must be considered. As example, we consider three levels in $^{85}$Rb: $5{}^{2}{{S}_{1/2}}$ as $\left| c \right\rangle $,  $8{}^{2}{{S}_{1/2}}$ as $\left| a \right\rangle $, and $8{}^{2}{{P}_{3/2}}$ as $\left| b \right\rangle $. The transition wavelengths ${{\lambda }_{bc}}={2\pi c}/{{{\omega }_{bc}}}\;=335\text{nm}$ and ${{\lambda }_{ba}}={2\pi c}/{{{\omega }_{ba}}}\;=12.40\mu \text{m}$ and thus $\lambda_{ba}/\lambda_{bc}=37.01$. Thanks to the very large dipole transition matrix element between $8{}^{2}{{S}_{1/2}}$ and $8{}^{2}{{P}_{3/2}}$, ${{\mu }_{ab}}=36.123e{{a}_{0}}\approx 3.06\times {{10}^{-28}}$C$\cdot$m \cite{Safronova2004} where ${a}_{0}$ is the Bohr radius, a laser with intensity $I=39{\text{ }\!\!\mu\!\!\text{ W}}/{\text{m}{{\text{m}}^{\text{2}}}}\;$ can induce a Rabi frequency as large as $\Omega_1=4\times 10^7\sqrt{I\left(\mu\text{W}/\text{mm}^2\right)}=2.5\times {{10}^{8}}{{s}^{-1}}=200{{\gamma }_{bc}}$.

\begin{figure}[t]
\begin{center}
\epsfig{figure=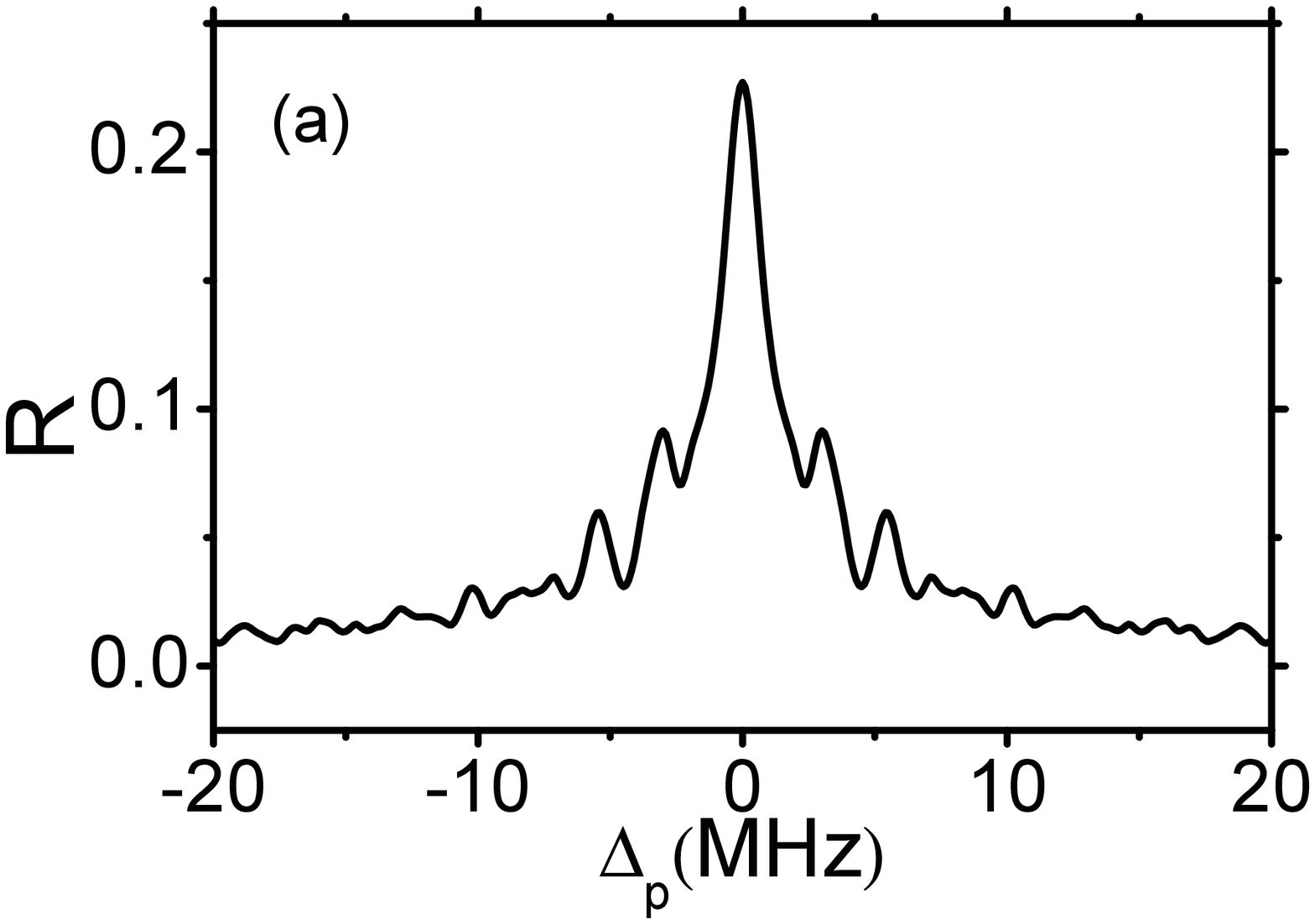, angle=0, width=0.23\textwidth}
\epsfig{figure=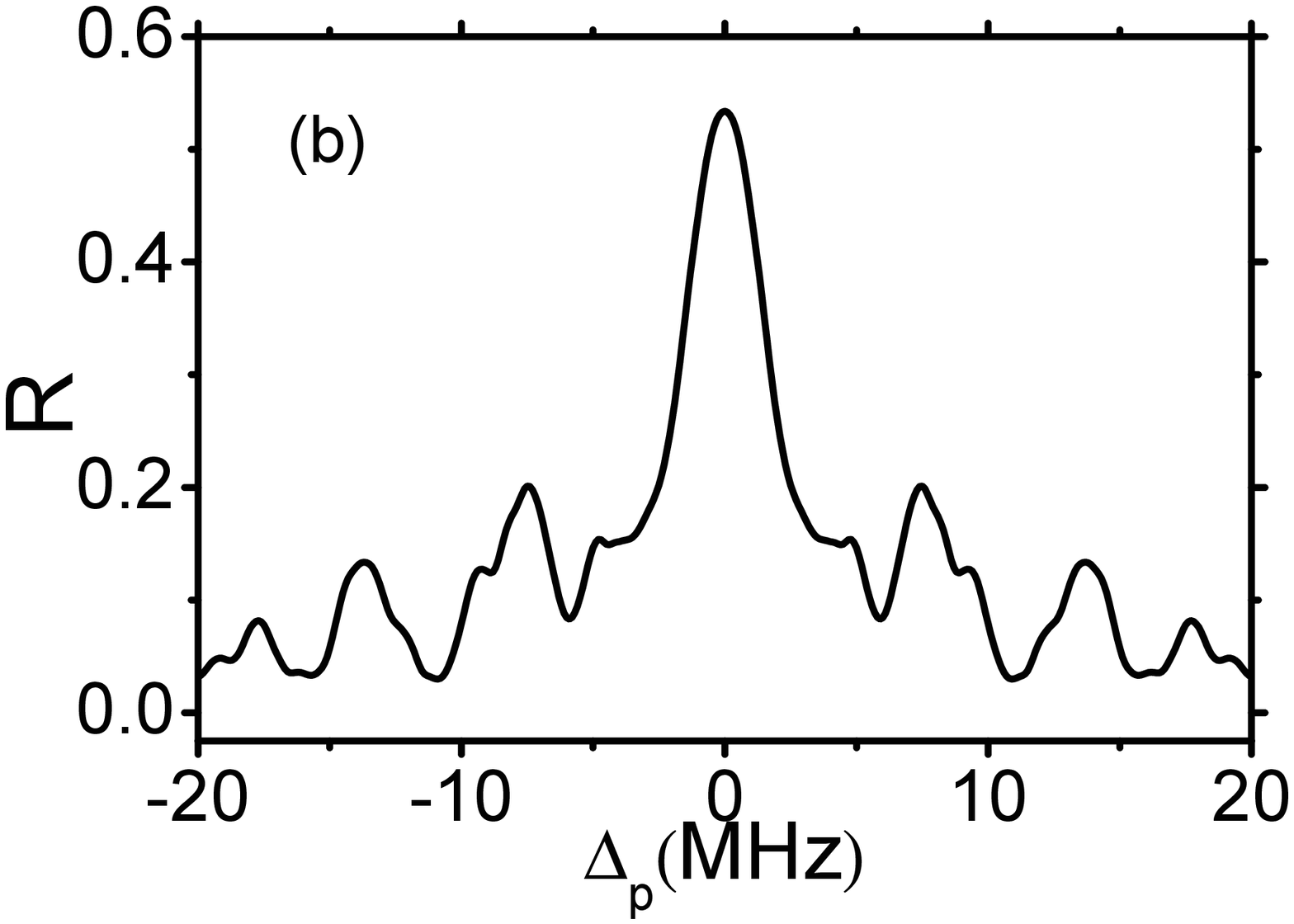, angle=0, width=0.23\textwidth}
\caption{The reflectivity from the 37th order photonic band gap in thermal $^{85}$Rb vapor at 485K. ${{\gamma }_{bc}}=1.25\times {{10}^{6}}{{\text{s}}^{-1}}$, ${\gamma }_{ac}=3.25\times {{10}^{6}}{{\text{s}}^{-1}}$ \cite{Theodosiou1984}, $\Delta_1=0$. $\lambda_{ba}=37.01\lambda_{bc}$=12.4$\mu$m, $\theta=0.0232$ radian so that $\Delta k_{n}^{0}=0$. The sample length is $10^{5}\lambda_{bc}$=3.36cm. (a) ${{\Omega }_{1}}={{\Omega }_{2}}=200{{\gamma }_{bc}}$; (b) ${{\Omega }_{1}}={{\Omega }_{2}}=500{{\gamma }_{bc}}$.}
\label{bg_hot}
\end{center}
\end{figure}

In Fig.~\ref{bg_hot}, we plot the reflectance in a thermal $^{85}$Rb vapor cell at 485K. Although the dephasing rate is $\gamma_{ac}=2.6\gamma_{bc}=3.25\times {{10}^{6}}{{\text{s}}^{-1}}$ and the Doppler broadening is 696MHz, the reflectivity from the 37th order band gap or 76-wave-mixing can still exceed 50\% for $\Omega_1=500\gamma_{bc}$ (which only requires a moderate driving laser intensity 244$\mu\text{W/mm}^2$). In calculating the susceptibilities including the movement of the atoms, we use the technique of continued fractions and average over the Maxwellian velocity distribution \cite{Zhang2011a}. The main effect of $\gamma_{ac}$ is the incoherent absorption of the probe light during which the energy is dissipated by transitions into other levels. A ladder system with $\left|a\right\rangle$ higher than $\left|b\right\rangle$ works equally well or even better if the life time of $\left|a\right\rangle$ is longer (which is true for Rydberg states). By increasing $\Omega_1$ from $200\gamma_{bc}$ to $500\gamma_{bc}$, the atom remains shorter in  $\left|a\right\rangle$  before the coherent backward photon is emitted and dissipation is suppressed, increasing the reflectivity from 0.2 to 0.5.

\begin{figure}[t]
\begin{center}
\epsfig{figure=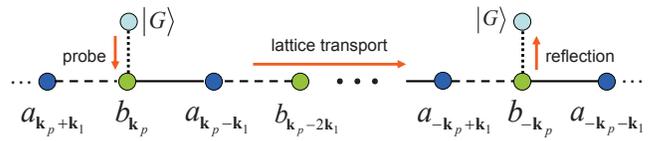, angle=0, width=0.48\textwidth}
\end{center}
\caption{(Color online) Mechanism of the reflection in the picture of the superradiance lattice. Solid (dash) lines denote the coupling field $\Omega_1$ ($\Omega_2$). Dotted line denotes the superradiance coupling between $|b_{\pm \mathbf{k}_p}\ra$ and the ground state $|G\ra$ (gray circle). Blue (green) circles denote states $a$ ($b$). The red arrows show the direction of the dynamic evolution of the excitation which results in the reflection.}
\label{lattice}
\end{figure}

The physics of the reflection can be understood from the picture of the superradiance lattice \cite{Wang2015}, as shown in Fig. \ref{lattice}. The probe field creates excitation $|b_{\mathbf{k}_p}\ra$ from the ground state $|G\ra\equiv |c_1, c_2,...,c_N\ra$ in a one-dimensional momentum space tight-binding lattice where the lattice sites are timed Dicke states connected by the two coupling fields $\Omega_1$ and $\Omega_2$ \cite{Wang2015}. The excitation propagates along the lattice to $|b_{-\mathbf{k}_p}\ra$ which is strongly coupled to the ground state $|G\ra$ due to the superradiance $\sqrt{N}$ enhancement, and consequently generates the reflected field. The other states between or outside of $|b_{\mathbf{k}_p}\ra$ and $|b_{-\mathbf{k}_p}\ra$ are only weakly coupled to the ground state $|G\ra$ via the two endfire modes $\pm\mathbf{k}_p$ due to the subradiant effect. This is therefore a Fano lattice \cite{Miroshnichenko2010} in momentum space, which explains the Bessel function in Eq. (\ref{p2n}) and the Fano feature in Fig. \ref{bg_cold}. The superradiance lattice gives the continuum and the leakage from $|b_{-\mathbf{k}_p}\ra$ to $|G\ra$ gives the discrete channel for Fano resonances \cite{Fano1961}. The nonlinearity is governed by lattice dynamics and results in the linear scaling in Eq. (\ref{criteria}) rather than the power law dependence in conventional nonlinear optics.

X-ray EIT has been proposed with inner shell transitions in gases \cite{Buth2007}, and some optical control of X-ray transmission was realized~\cite{Glover2010}. Recently, X-ray frequency combs are proposed based on a three-level configuration of Be$^{2+}$ ions \cite{Cavaletto2014}. Our scheme can be directly applied to the same energy levels, namely, $c=1s^2{^1}S_0$, $b=1s2p ^1P_1$ and $a=1s2s{^1}S_0$. The decoherence rates are $\gamma_{bc}=\Gamma_{bc}/2=6\times10^{10}s^{-1}$ and $\gamma_{ac}=9\times 10^3s^{-1}$ which is negligible. The transition energies are $\hbar\omega_{bc}=123.7$eV (10nm) and $\hbar\omega_{ba}=2.02$eV (614nm). We can use the 61st order photonic band gap with an incidence angle near $\theta=0.088$. The Rabi frequency is $\Omega_{1,2}>n\gamma_{bc}=3.6\times 10^{12}s^{-1}$. The intensity required is only in the order of $10^{9}$W/cm$^2$ and safe for ionization. Present challenges for the experimental implementation are the relatively high temperature and low density of the ions, which however, can be overcome by the cooling techniques recently developed \cite{Hansen2014}.

For hard X-ray, EIT has been experimentally demonstrated with the 14.4~keV nuclear M\"o{\ss}bauer transition in $^{57}$Fe~\cite{Rohlsberger2012}. Multi-level schemes can also be engineered which could be driven by multiple incident fields, and which are essentially decoherence free~\cite{Heeg2013}. This together with the rapid development~\cite{Coussement2002,Gheysen2006,Rohlsberger2010} in this field and the upcoming availability of temporally coherent X-ray free electron lasers in the hard X-ray regime renders nuclear quantum optics a promising platform to realize HOPBGs via EIT.

In conclusion, the reflection of high frequency light from the spatial coherence generated by low frequency light is studied. The possibility of the light-controllable photonic band gaps was mentioned in a paper on reflection combs \cite{Raczynski2009}. Here we analysed the scaling of the band gaps on high order $n$. The required driving field strength scales linearly with $n$ in contrast with the power law scaling in conventional nonlinear optics. The physics is envisioned by superradiance lattices. Experiments can be done in Rb atoms (infrared reflects ultraviolet) or Be$^{2+}$ ions (visible light reflects soft X-ray).

We thank Ren-Bao Liu for helpful discussion. We gratefully acknowledge the support of the National Science Foundation Grants No. PHY-1241032(INSPIRE CREATIV) and PHY-1068554 and the Robert A. Welch Foundation (Grant No. A-1261). S.Y.Z. was suported by National Basic Research Program of China No. 2012CB921603 and National Natural Science Foundation of China No. U1330203.

\bibliographystyle{apsrev4-1}
\bibliography{hopbg}

\begin{thebibliography}{35}%
\makeatletter
\providecommand \@ifxundefined [1]{%
 \@ifx{#1\undefined}
}%
\providecommand \@ifnum [1]{%
 \ifnum #1\expandafter \@firstoftwo
 \else \expandafter \@secondoftwo
 \fi
}%
\providecommand \@ifx [1]{%
 \ifx #1\expandafter \@firstoftwo
 \else \expandafter \@secondoftwo
 \fi
}%
\providecommand \natexlab [1]{#1}%
\providecommand \enquote  [1]{``#1''}%
\providecommand \bibnamefont  [1]{#1}%
\providecommand \bibfnamefont [1]{#1}%
\providecommand \citenamefont [1]{#1}%
\providecommand \href@noop [0]{\@secondoftwo}%
\providecommand \href [0]{\begingroup \@sanitize@url \@href}%
\providecommand \@href[1]{\@@startlink{#1}\@@href}%
\providecommand \@@href[1]{\endgroup#1\@@endlink}%
\providecommand \@sanitize@url [0]{\catcode `\\12\catcode `\$12\catcode
  `\&12\catcode `\#12\catcode `\^12\catcode `\_12\catcode `\%12\relax}%
\providecommand \@@startlink[1]{}%
\providecommand \@@endlink[0]{}%
\providecommand \url  [0]{\begingroup\@sanitize@url \@url }%
\providecommand \@url [1]{\endgroup\@href {#1}{\urlprefix }}%
\providecommand \urlprefix  [0]{URL }%
\providecommand \Eprint [0]{\href }%
\providecommand \doibase [0]{http://dx.doi.org/}%
\providecommand \selectlanguage [0]{\@gobble}%
\providecommand \bibinfo  [0]{\@secondoftwo}%
\providecommand \bibfield  [0]{\@secondoftwo}%
\providecommand \translation [1]{[#1]}%
\providecommand \BibitemOpen [0]{}%
\providecommand \bibitemStop [0]{}%
\providecommand \bibitemNoStop [0]{.\EOS\space}%
\providecommand \EOS [0]{\spacefactor3000\relax}%
\providecommand \BibitemShut  [1]{\csname bibitem#1\endcsname}%
\let\auto@bib@innerbib\@empty
\bibitem [{\citenamefont {Yablonovitch}(1987)}]{Yablonovitch1987}%
  \BibitemOpen
  \bibfield  {author} {\bibinfo {author} {\bibfnamefont {E.}~\bibnamefont
  {Yablonovitch}},\ }\href@noop {} {\bibfield  {journal} {\bibinfo  {journal}
  {Physical Review Letters}\ }\textbf {\bibinfo {volume} {58}},\ \bibinfo
  {pages} {2059} (\bibinfo {year} {1987})}\BibitemShut {NoStop}%
\bibitem [{\citenamefont {John}(1987)}]{John1987}%
  \BibitemOpen
  \bibfield  {author} {\bibinfo {author} {\bibfnamefont {S.}~\bibnamefont
  {John}},\ }\href@noop {} {\bibfield  {journal} {\bibinfo  {journal} {Physical
  Review Letters}\ }\textbf {\bibinfo {volume} {58}},\ \bibinfo {pages} {2486}
  (\bibinfo {year} {1987})}\BibitemShut {NoStop}%
\bibitem [{\citenamefont {Wu}\ \emph {et~al.}(2004)\citenamefont {Wu},
  \citenamefont {Yamilov}, \citenamefont {Liu}, \citenamefont {Li},
  \citenamefont {Dravid}, \citenamefont {Chang},\ and\ \citenamefont
  {Cao}}]{Wu2004}%
  \BibitemOpen
  \bibfield  {author} {\bibinfo {author} {\bibfnamefont {X.}~\bibnamefont
  {Wu}}, \bibinfo {author} {\bibfnamefont {A.}~\bibnamefont {Yamilov}},
  \bibinfo {author} {\bibfnamefont {X.}~\bibnamefont {Liu}}, \bibinfo {author}
  {\bibfnamefont {S.}~\bibnamefont {Li}}, \bibinfo {author} {\bibfnamefont
  {V.~P.}\ \bibnamefont {Dravid}}, \bibinfo {author} {\bibfnamefont {R.~P.~H.}\
  \bibnamefont {Chang}}, \ and\ \bibinfo {author} {\bibfnamefont
  {H.}~\bibnamefont {Cao}},\ }\href@noop {} {\bibfield  {journal} {\bibinfo
  {journal} {Applied Physics Letters}\ }\textbf {\bibinfo {volume} {85}},\
  \bibinfo {pages} {3657} (\bibinfo {year} {2004})}\BibitemShut {NoStop}%
\bibitem [{\citenamefont {Ganesh}\ and\ \citenamefont
  {Cunningham}(2006)}]{Ganesh2006}%
  \BibitemOpen
  \bibfield  {author} {\bibinfo {author} {\bibfnamefont {N.}~\bibnamefont
  {Ganesh}}\ and\ \bibinfo {author} {\bibfnamefont {B.~T.}\ \bibnamefont
  {Cunningham}},\ }\href@noop {} {\bibfield  {journal} {\bibinfo  {journal}
  {Applied Physics Letters}\ }\textbf {\bibinfo {volume} {88}},\ \bibinfo
  {pages} {071110} (\bibinfo {year} {2006})}\BibitemShut {NoStop}%
\bibitem [{\citenamefont {Radeonychev}\ \emph {et~al.}(2009)\citenamefont
  {Radeonychev}, \citenamefont {Koryukin},\ and\ \citenamefont
  {Kocharovskaya}}]{Radeonychev2009}%
  \BibitemOpen
  \bibfield  {author} {\bibinfo {author} {\bibfnamefont {Y.~V.}\ \bibnamefont
  {Radeonychev}}, \bibinfo {author} {\bibfnamefont {I.~V.}\ \bibnamefont
  {Koryukin}}, \ and\ \bibinfo {author} {\bibfnamefont {O.}~\bibnamefont
  {Kocharovskaya}},\ }\href@noop {} {\bibfield  {journal} {\bibinfo  {journal}
  {Laser Physics}\ }\textbf {\bibinfo {volume} {19}},\ \bibinfo {pages} {1207}
  (\bibinfo {year} {2009})}\BibitemShut {NoStop}%
\bibitem [{\citenamefont {Shvyd'ko}\ \emph {et~al.}(2011)\citenamefont
  {Shvyd'ko}, \citenamefont {Stoupin}, \citenamefont {Blank},\ and\
  \citenamefont {Terentyev}}]{Shvydko2011}%
  \BibitemOpen
  \bibfield  {author} {\bibinfo {author} {\bibfnamefont {Y.}~\bibnamefont
  {Shvyd'ko}}, \bibinfo {author} {\bibfnamefont {S.}~\bibnamefont {Stoupin}},
  \bibinfo {author} {\bibfnamefont {V.}~\bibnamefont {Blank}}, \ and\ \bibinfo
  {author} {\bibfnamefont {S.}~\bibnamefont {Terentyev}},\ }\href@noop {}
  {\bibfield  {journal} {\bibinfo  {journal} {Nature Photonics}\ }\textbf
  {\bibinfo {volume} {5}},\ \bibinfo {pages} {539} (\bibinfo {year}
  {2011})}\BibitemShut {NoStop}%
\bibitem [{\citenamefont {Boller}\ \emph {et~al.}(1991)\citenamefont {Boller},
  \citenamefont {Imamoglu},\ and\ \citenamefont {Harris}}]{Boller1991}%
  \BibitemOpen
  \bibfield  {author} {\bibinfo {author} {\bibfnamefont {K.~J.}\ \bibnamefont
  {Boller}}, \bibinfo {author} {\bibfnamefont {A.}~\bibnamefont {Imamoglu}}, \
  and\ \bibinfo {author} {\bibfnamefont {S.~E.}\ \bibnamefont {Harris}},\
  }\href@noop {} {\bibfield  {journal} {\bibinfo  {journal} {Physical Review
  Letters}\ }\textbf {\bibinfo {volume} {66}},\ \bibinfo {pages} {2593}
  (\bibinfo {year} {1991})}\BibitemShut {NoStop}%
\bibitem [{\citenamefont {Andre}\ and\ \citenamefont
  {Lukin}(2002)}]{Andre2002}%
  \BibitemOpen
  \bibfield  {author} {\bibinfo {author} {\bibfnamefont {A.}~\bibnamefont
  {Andre}}\ and\ \bibinfo {author} {\bibfnamefont {M.~D.}\ \bibnamefont
  {Lukin}},\ }\href@noop {} {\bibfield  {journal} {\bibinfo  {journal}
  {Physical Review Letters}\ }\textbf {\bibinfo {volume} {89}},\ \bibinfo
  {pages} {143602} (\bibinfo {year} {2002})}\BibitemShut {NoStop}%
\bibitem [{\citenamefont {Artoni}\ and\ \citenamefont
  {La~Rocca}(2006)}]{Artoni2006}%
  \BibitemOpen
  \bibfield  {author} {\bibinfo {author} {\bibfnamefont {M.}~\bibnamefont
  {Artoni}}\ and\ \bibinfo {author} {\bibfnamefont {G.~C.}\ \bibnamefont
  {La~Rocca}},\ }\href@noop {} {\bibfield  {journal} {\bibinfo  {journal}
  {Physical Review Letters}\ }\textbf {\bibinfo {volume} {96}},\ \bibinfo
  {pages} {073905} (\bibinfo {year} {2006})}\BibitemShut {NoStop}%
\bibitem [{\citenamefont {Ranitovic}\ \emph {et~al.}(2011)\citenamefont
  {Ranitovic}, \citenamefont {Tong}, \citenamefont {Hogle}, \citenamefont
  {Zhou}, \citenamefont {Liu}, \citenamefont {Toshima}, \citenamefont
  {Murnane},\ and\ \citenamefont {Kapteyn}}]{Ranitovic2011}%
  \BibitemOpen
  \bibfield  {author} {\bibinfo {author} {\bibfnamefont {P.}~\bibnamefont
  {Ranitovic}}, \bibinfo {author} {\bibfnamefont {X.~M.}\ \bibnamefont {Tong}},
  \bibinfo {author} {\bibfnamefont {C.~W.}\ \bibnamefont {Hogle}}, \bibinfo
  {author} {\bibfnamefont {X.}~\bibnamefont {Zhou}}, \bibinfo {author}
  {\bibfnamefont {Y.}~\bibnamefont {Liu}}, \bibinfo {author} {\bibfnamefont
  {N.}~\bibnamefont {Toshima}}, \bibinfo {author} {\bibfnamefont {M.~M.}\
  \bibnamefont {Murnane}}, \ and\ \bibinfo {author} {\bibfnamefont {H.~C.}\
  \bibnamefont {Kapteyn}},\ }\href@noop {} {\bibfield  {journal} {\bibinfo
  {journal} {Physical Review Letters}\ }\textbf {\bibinfo {volume} {106}},\
  \bibinfo {pages} {193008} (\bibinfo {year} {2011})}\BibitemShut {NoStop}%
\bibitem [{\citenamefont {Glover}\ \emph {et~al.}(2010)\citenamefont {Glover},
  \citenamefont {Hertlein}, \citenamefont {Southworth}, \citenamefont
  {Allison}, \citenamefont {van Tilborg}, \citenamefont {Kanter}, \citenamefont
  {Krassig}, \citenamefont {Varma}, \citenamefont {Rude}, \citenamefont
  {Santra}, \citenamefont {Belkacem},\ and\ \citenamefont
  {Young}}]{Glover2010}%
  \BibitemOpen
  \bibfield  {author} {\bibinfo {author} {\bibfnamefont {T.~E.}\ \bibnamefont
  {Glover}}, \bibinfo {author} {\bibfnamefont {M.~P.}\ \bibnamefont
  {Hertlein}}, \bibinfo {author} {\bibfnamefont {S.~H.}\ \bibnamefont
  {Southworth}}, \bibinfo {author} {\bibfnamefont {T.~K.}\ \bibnamefont
  {Allison}}, \bibinfo {author} {\bibfnamefont {J.}~\bibnamefont {van
  Tilborg}}, \bibinfo {author} {\bibfnamefont {E.~P.}\ \bibnamefont {Kanter}},
  \bibinfo {author} {\bibfnamefont {B.}~\bibnamefont {Krassig}}, \bibinfo
  {author} {\bibfnamefont {H.~R.}\ \bibnamefont {Varma}}, \bibinfo {author}
  {\bibfnamefont {B.}~\bibnamefont {Rude}}, \bibinfo {author} {\bibfnamefont
  {R.}~\bibnamefont {Santra}}, \bibinfo {author} {\bibfnamefont
  {A.}~\bibnamefont {Belkacem}}, \ and\ \bibinfo {author} {\bibfnamefont
  {L.}~\bibnamefont {Young}},\ }\href@noop {} {\bibfield  {journal} {\bibinfo
  {journal} {Nat Phys}\ }\textbf {\bibinfo {volume} {6}},\ \bibinfo {pages}
  {69} (\bibinfo {year} {2010})}\BibitemShut {NoStop}%
\bibitem [{\citenamefont {Straub}\ \emph {et~al.}(2003)\citenamefont {Straub},
  \citenamefont {Ventura},\ and\ \citenamefont {Gu}}]{Straub2003}%
  \BibitemOpen
  \bibfield  {author} {\bibinfo {author} {\bibfnamefont {M.}~\bibnamefont
  {Straub}}, \bibinfo {author} {\bibfnamefont {M.}~\bibnamefont {Ventura}}, \
  and\ \bibinfo {author} {\bibfnamefont {M.}~\bibnamefont {Gu}},\ }\href@noop
  {} {\bibfield  {journal} {\bibinfo  {journal} {Physical Review Letters}\
  }\textbf {\bibinfo {volume} {91}},\ \bibinfo {pages} {043901} (\bibinfo
  {year} {2003})}\BibitemShut {NoStop}%
\bibitem [{\citenamefont {Barillaro}\ \emph {et~al.}(2007)\citenamefont
  {Barillaro}, \citenamefont {Annovazzi-Lodi}, \citenamefont {Benedetti},\ and\
  \citenamefont {Merlo}}]{Barillaro2007}%
  \BibitemOpen
  \bibfield  {author} {\bibinfo {author} {\bibfnamefont {G.}~\bibnamefont
  {Barillaro}}, \bibinfo {author} {\bibfnamefont {V.}~\bibnamefont
  {Annovazzi-Lodi}}, \bibinfo {author} {\bibfnamefont {M.}~\bibnamefont
  {Benedetti}}, \ and\ \bibinfo {author} {\bibfnamefont {S.}~\bibnamefont
  {Merlo}},\ }\href@noop {} {\bibfield  {journal} {\bibinfo  {journal} {Applied
  Physics Letters}\ }\textbf {\bibinfo {volume} {90}} (\bibinfo {year}
  {2007})}\BibitemShut {NoStop}%
\bibitem [{\citenamefont {Morozov}\ and\ \citenamefont
  {Placido}(2010)}]{Morozov2010}%
  \BibitemOpen
  \bibfield  {author} {\bibinfo {author} {\bibfnamefont {G.~V.}\ \bibnamefont
  {Morozov}}\ and\ \bibinfo {author} {\bibfnamefont {F.}~\bibnamefont
  {Placido}},\ }\href@noop {} {\bibfield  {journal} {\bibinfo  {journal}
  {Journal of Optics}\ }\textbf {\bibinfo {volume} {12}},\ \bibinfo {pages}
  {045101} (\bibinfo {year} {2010})}\BibitemShut {NoStop}%
\bibitem [{\citenamefont {Lu}\ \emph {et~al.}(2012)\citenamefont {Lu},
  \citenamefont {Chi}, \citenamefont {Zhou},\ and\ \citenamefont
  {Lun}}]{Lu2012}%
  \BibitemOpen
  \bibfield  {author} {\bibinfo {author} {\bibfnamefont {X.}~\bibnamefont
  {Lu}}, \bibinfo {author} {\bibfnamefont {F.}~\bibnamefont {Chi}}, \bibinfo
  {author} {\bibfnamefont {T.}~\bibnamefont {Zhou}}, \ and\ \bibinfo {author}
  {\bibfnamefont {S.}~\bibnamefont {Lun}},\ }\href@noop {} {\bibfield
  {journal} {\bibinfo  {journal} {Optics Communications}\ }\textbf {\bibinfo
  {volume} {285}},\ \bibinfo {pages} {1885} (\bibinfo {year}
  {2012})}\BibitemShut {NoStop}%
\bibitem [{\citenamefont {Miroshnichenko}\ \emph {et~al.}(2010)\citenamefont
  {Miroshnichenko}, \citenamefont {Flach},\ and\ \citenamefont
  {Kivshar}}]{Miroshnichenko2010}%
  \BibitemOpen
  \bibfield  {author} {\bibinfo {author} {\bibfnamefont {A.~E.}\ \bibnamefont
  {Miroshnichenko}}, \bibinfo {author} {\bibfnamefont {S.}~\bibnamefont
  {Flach}}, \ and\ \bibinfo {author} {\bibfnamefont {Y.~S.}\ \bibnamefont
  {Kivshar}},\ }\href@noop {} {\bibfield  {journal} {\bibinfo  {journal}
  {Reviews of Modern Physics}\ }\textbf {\bibinfo {volume} {82}},\ \bibinfo
  {pages} {2257} (\bibinfo {year} {2010})}\BibitemShut {NoStop}%
\bibitem [{\citenamefont {Wang}\ \emph {et~al.}(2015)\citenamefont {Wang},
  \citenamefont {Liu}, \citenamefont {Zhu},\ and\ \citenamefont
  {Scully}}]{Wang2015}%
  \BibitemOpen
  \bibfield  {author} {\bibinfo {author} {\bibfnamefont {D.-W.}\ \bibnamefont
  {Wang}}, \bibinfo {author} {\bibfnamefont {R.-B.}\ \bibnamefont {Liu}},
  \bibinfo {author} {\bibfnamefont {S.-Y.}\ \bibnamefont {Zhu}}, \ and\
  \bibinfo {author} {\bibfnamefont {M.~O.}\ \bibnamefont {Scully}},\
  }\href@noop {} {\bibfield  {journal} {\bibinfo  {journal} {Physical Review
  Letters}\ }\textbf {\bibinfo {volume} {114}},\ \bibinfo {pages} {043602}
  (\bibinfo {year} {2015})}\BibitemShut {NoStop}%
\bibitem [{\citenamefont {Scully}\ \emph {et~al.}(2006)\citenamefont {Scully},
  \citenamefont {Fry}, \citenamefont {Ooi},\ and\ \citenamefont
  {Wodkiewicz}}]{Scully2006}%
  \BibitemOpen
  \bibfield  {author} {\bibinfo {author} {\bibfnamefont {M.~O.}\ \bibnamefont
  {Scully}}, \bibinfo {author} {\bibfnamefont {E.~S.}\ \bibnamefont {Fry}},
  \bibinfo {author} {\bibfnamefont {C.~H.~R.}\ \bibnamefont {Ooi}}, \ and\
  \bibinfo {author} {\bibfnamefont {K.}~\bibnamefont {Wodkiewicz}},\
  }\href@noop {} {\bibfield  {journal} {\bibinfo  {journal} {Physical Review
  Letters}\ }\textbf {\bibinfo {volume} {96}},\ \bibinfo {pages} {010501}
  (\bibinfo {year} {2006})}\BibitemShut {NoStop}%
\bibitem [{\citenamefont {Liao}\ \emph {et~al.}(2010)\citenamefont {Liao},
  \citenamefont {Al-Amri},\ and\ \citenamefont {Suhail~Zubairy}}]{Liao2010}%
  \BibitemOpen
  \bibfield  {author} {\bibinfo {author} {\bibfnamefont {Z.}~\bibnamefont
  {Liao}}, \bibinfo {author} {\bibfnamefont {M.}~\bibnamefont {Al-Amri}}, \
  and\ \bibinfo {author} {\bibfnamefont {M.}~\bibnamefont {Suhail~Zubairy}},\
  }\href@noop {} {\bibfield  {journal} {\bibinfo  {journal} {Physical Review
  Letters}\ }\textbf {\bibinfo {volume} {105}},\ \bibinfo {pages} {183601}
  (\bibinfo {year} {2010})}\BibitemShut {NoStop}%
\bibitem [{\citenamefont {Wang}\ and\ \citenamefont
  {Scully}(2014)}]{Wang2014a}%
  \BibitemOpen
  \bibfield  {author} {\bibinfo {author} {\bibfnamefont {D.-W.}\ \bibnamefont
  {Wang}}\ and\ \bibinfo {author} {\bibfnamefont {M.~O.}\ \bibnamefont
  {Scully}},\ }\href@noop {} {\bibfield  {journal} {\bibinfo  {journal}
  {Physical Review Letters}\ }\textbf {\bibinfo {volume} {113}},\ \bibinfo
  {pages} {083601} (\bibinfo {year} {2014})}\BibitemShut {NoStop}%
\bibitem [{\citenamefont {Wang}\ \emph {et~al.}(2013)\citenamefont {Wang},
  \citenamefont {Zhou}, \citenamefont {Guo}, \citenamefont {Zhang},
  \citenamefont {Evers},\ and\ \citenamefont {Zhu}}]{Wang2013}%
  \BibitemOpen
  \bibfield  {author} {\bibinfo {author} {\bibfnamefont {D.-W.}\ \bibnamefont
  {Wang}}, \bibinfo {author} {\bibfnamefont {H.-T.}\ \bibnamefont {Zhou}},
  \bibinfo {author} {\bibfnamefont {M.-J.}\ \bibnamefont {Guo}}, \bibinfo
  {author} {\bibfnamefont {J.-X.}\ \bibnamefont {Zhang}}, \bibinfo {author}
  {\bibfnamefont {J.}~\bibnamefont {Evers}}, \ and\ \bibinfo {author}
  {\bibfnamefont {S.-Y.}\ \bibnamefont {Zhu}},\ }\href@noop {} {\bibfield
  {journal} {\bibinfo  {journal} {Physical Review Letters}\ }\textbf {\bibinfo
  {volume} {110}},\ \bibinfo {pages} {093901} (\bibinfo {year}
  {2013})}\BibitemShut {NoStop}%
\bibitem [{\citenamefont {Wang}(2014)}]{Wang2014}%
  \BibitemOpen
  \bibfield  {author} {\bibinfo {author} {\bibfnamefont {D.}~\bibnamefont
  {Wang}},\ }\href@noop {} {\emph {\bibinfo {title} {Atom-photon Interactions
  without RWA and Standing Wave Coupled EIT: Virtual processes, quantum
  interference and their applications}}}\ (\bibinfo  {publisher} {PhD thesis
  (2012), LAP},\ \bibinfo {year} {2014})\BibitemShut {NoStop}%
\bibitem [{\citenamefont {Safronova}\ \emph {et~al.}(2004)\citenamefont
  {Safronova}, \citenamefont {Williams},\ and\ \citenamefont
  {Clark}}]{Safronova2004}%
  \BibitemOpen
  \bibfield  {author} {\bibinfo {author} {\bibfnamefont {M.~S.}\ \bibnamefont
  {Safronova}}, \bibinfo {author} {\bibfnamefont {C.~J.}\ \bibnamefont
  {Williams}}, \ and\ \bibinfo {author} {\bibfnamefont {C.~W.}\ \bibnamefont
  {Clark}},\ }\href@noop {} {\bibfield  {journal} {\bibinfo  {journal}
  {Physical Review A}\ }\textbf {\bibinfo {volume} {69}},\ \bibinfo {pages}
  {022509} (\bibinfo {year} {2004})}\BibitemShut {NoStop}%
\bibitem [{\citenamefont {Zhang}\ \emph {et~al.}(2011)\citenamefont {Zhang},
  \citenamefont {Zhou}, \citenamefont {Wang},\ and\ \citenamefont
  {Zhu}}]{Zhang2011a}%
  \BibitemOpen
  \bibfield  {author} {\bibinfo {author} {\bibfnamefont {J.~X.}\ \bibnamefont
  {Zhang}}, \bibinfo {author} {\bibfnamefont {H.~T.}\ \bibnamefont {Zhou}},
  \bibinfo {author} {\bibfnamefont {D.~W.}\ \bibnamefont {Wang}}, \ and\
  \bibinfo {author} {\bibfnamefont {S.~Y.}\ \bibnamefont {Zhu}},\ }\href@noop
  {} {\bibfield  {journal} {\bibinfo  {journal} {Physical Review A}\ }\textbf
  {\bibinfo {volume} {83}},\ \bibinfo {pages} {053841} (\bibinfo {year}
  {2011})}\BibitemShut {NoStop}%
\bibitem [{\citenamefont {Theodosiou}(1984)}]{Theodosiou1984}%
  \BibitemOpen
  \bibfield  {author} {\bibinfo {author} {\bibfnamefont {C.~E.}\ \bibnamefont
  {Theodosiou}},\ }\href@noop {} {\bibfield  {journal} {\bibinfo  {journal}
  {Physical Review A}\ }\textbf {\bibinfo {volume} {30}},\ \bibinfo {pages}
  {2881} (\bibinfo {year} {1984})}\BibitemShut {NoStop}%
\bibitem [{\citenamefont {Fano}(1961)}]{Fano1961}%
  \BibitemOpen
  \bibfield  {author} {\bibinfo {author} {\bibfnamefont {U.}~\bibnamefont
  {Fano}},\ }\href@noop {} {\bibfield  {journal} {\bibinfo  {journal} {Physical
  Review}\ }\textbf {\bibinfo {volume} {124}},\ \bibinfo {pages} {1866}
  (\bibinfo {year} {1961})}\BibitemShut {NoStop}%
\bibitem [{\citenamefont {Buth}\ \emph {et~al.}(2007)\citenamefont {Buth},
  \citenamefont {Santra},\ and\ \citenamefont {Young}}]{Buth2007}%
  \BibitemOpen
  \bibfield  {author} {\bibinfo {author} {\bibfnamefont {C.}~\bibnamefont
  {Buth}}, \bibinfo {author} {\bibfnamefont {R.}~\bibnamefont {Santra}}, \ and\
  \bibinfo {author} {\bibfnamefont {L.}~\bibnamefont {Young}},\ }\href@noop {}
  {\bibfield  {journal} {\bibinfo  {journal} {Physical Review Letters}\
  }\textbf {\bibinfo {volume} {98}},\ \bibinfo {pages} {253001} (\bibinfo
  {year} {2007})}\BibitemShut {NoStop}%
\bibitem [{\citenamefont {Cavaletto}\ \emph {et~al.}(2014)\citenamefont
  {Cavaletto}, \citenamefont {Harman}, \citenamefont {Ott}, \citenamefont
  {Buth}, \citenamefont {Pfeifer},\ and\ \citenamefont
  {Keitel}}]{Cavaletto2014}%
  \BibitemOpen
  \bibfield  {author} {\bibinfo {author} {\bibfnamefont {S.~M.}\ \bibnamefont
  {Cavaletto}}, \bibinfo {author} {\bibfnamefont {Z.}~\bibnamefont {Harman}},
  \bibinfo {author} {\bibfnamefont {C.}~\bibnamefont {Ott}}, \bibinfo {author}
  {\bibfnamefont {C.}~\bibnamefont {Buth}}, \bibinfo {author} {\bibfnamefont
  {T.}~\bibnamefont {Pfeifer}}, \ and\ \bibinfo {author} {\bibfnamefont
  {C.~H.}\ \bibnamefont {Keitel}},\ }\href@noop {} {\bibfield  {journal}
  {\bibinfo  {journal} {Nat Photon}\ }\textbf {\bibinfo {volume} {8}},\
  \bibinfo {pages} {520} (\bibinfo {year} {2014})}\BibitemShut {NoStop}%
\bibitem [{\citenamefont {Hansen}\ \emph {et~al.}(2014)\citenamefont {Hansen},
  \citenamefont {Versolato}, \citenamefont {Klosowski}, \citenamefont
  {Kristensen}, \citenamefont {Gingell}, \citenamefont {Schwarz}, \citenamefont
  {Windberger}, \citenamefont {Ullrich}, \citenamefont {Lopez-Urrutia},\ and\
  \citenamefont {Drewsen}}]{Hansen2014}%
  \BibitemOpen
  \bibfield  {author} {\bibinfo {author} {\bibfnamefont {A.~K.}\ \bibnamefont
  {Hansen}}, \bibinfo {author} {\bibfnamefont {O.~O.}\ \bibnamefont
  {Versolato}}, \bibinfo {author} {\bibfnamefont {L.}~\bibnamefont
  {Klosowski}}, \bibinfo {author} {\bibfnamefont {S.~B.}\ \bibnamefont
  {Kristensen}}, \bibinfo {author} {\bibfnamefont {A.}~\bibnamefont {Gingell}},
  \bibinfo {author} {\bibfnamefont {M.}~\bibnamefont {Schwarz}}, \bibinfo
  {author} {\bibfnamefont {A.}~\bibnamefont {Windberger}}, \bibinfo {author}
  {\bibfnamefont {J.}~\bibnamefont {Ullrich}}, \bibinfo {author} {\bibfnamefont
  {J.~R.~C.}\ \bibnamefont {Lopez-Urrutia}}, \ and\ \bibinfo {author}
  {\bibfnamefont {M.}~\bibnamefont {Drewsen}},\ }\href@noop {} {\bibfield
  {journal} {\bibinfo  {journal} {Nature}\ }\textbf {\bibinfo {volume} {508}},\
  \bibinfo {pages} {76} (\bibinfo {year} {2014})}\BibitemShut {NoStop}%
\bibitem [{\citenamefont {R\"ohlsberger}\ \emph {et~al.}(2012)\citenamefont
  {R\"ohlsberger}, \citenamefont {Wille}, \citenamefont {Schlage},\ and\
  \citenamefont {Sahoo}}]{Rohlsberger2012}%
  \BibitemOpen
  \bibfield  {author} {\bibinfo {author} {\bibfnamefont {R.}~\bibnamefont
  {R\"ohlsberger}}, \bibinfo {author} {\bibfnamefont {H.-C.}\ \bibnamefont
  {Wille}}, \bibinfo {author} {\bibfnamefont {K.}~\bibnamefont {Schlage}}, \
  and\ \bibinfo {author} {\bibfnamefont {B.}~\bibnamefont {Sahoo}},\
  }\href@noop {} {\bibfield  {journal} {\bibinfo  {journal} {Nature}\ }\textbf
  {\bibinfo {volume} {482}},\ \bibinfo {pages} {199} (\bibinfo {year}
  {2012})}\BibitemShut {NoStop}%
\bibitem [{\citenamefont {Heeg}\ \emph {et~al.}(2013)\citenamefont {Heeg},
  \citenamefont {Wille}, \citenamefont {Schlage}, \citenamefont {Guryeva},
  \citenamefont {Schumacher}, \citenamefont {Uschmann}, \citenamefont
  {Schulze}, \citenamefont {Marx}, \citenamefont {Kämpfer}, \citenamefont
  {Paulus}, \citenamefont {Röhlsberger},\ and\ \citenamefont
  {Evers}}]{Heeg2013}%
  \BibitemOpen
  \bibfield  {author} {\bibinfo {author} {\bibfnamefont {K.~P.}\ \bibnamefont
  {Heeg}}, \bibinfo {author} {\bibfnamefont {H.-C.}\ \bibnamefont {Wille}},
  \bibinfo {author} {\bibfnamefont {K.}~\bibnamefont {Schlage}}, \bibinfo
  {author} {\bibfnamefont {T.}~\bibnamefont {Guryeva}}, \bibinfo {author}
  {\bibfnamefont {D.}~\bibnamefont {Schumacher}}, \bibinfo {author}
  {\bibfnamefont {I.}~\bibnamefont {Uschmann}}, \bibinfo {author}
  {\bibfnamefont {K.~S.}\ \bibnamefont {Schulze}}, \bibinfo {author}
  {\bibfnamefont {B.}~\bibnamefont {Marx}}, \bibinfo {author} {\bibfnamefont
  {T.}~\bibnamefont {Kämpfer}}, \bibinfo {author} {\bibfnamefont {G.~G.}\
  \bibnamefont {Paulus}}, \bibinfo {author} {\bibfnamefont {R.}~\bibnamefont
  {Röhlsberger}}, \ and\ \bibinfo {author} {\bibfnamefont {J.}~\bibnamefont
  {Evers}},\ }\href@noop {} {\bibfield  {journal} {\bibinfo  {journal}
  {Physical Review Letters}\ }\textbf {\bibinfo {volume} {111}},\ \bibinfo
  {pages} {073601} (\bibinfo {year} {2013})}\BibitemShut {NoStop}%
\bibitem [{\citenamefont {Coussement}\ \emph {et~al.}(2002)\citenamefont
  {Coussement}, \citenamefont {Rostovtsev}, \citenamefont {Odeurs},
  \citenamefont {Neyens}, \citenamefont {Muramatsu}, \citenamefont {Gheysen},
  \citenamefont {Callens}, \citenamefont {Vyvey}, \citenamefont {Kozyreff},
  \citenamefont {Mandel}, \citenamefont {Shakhmuratov},\ and\ \citenamefont
  {Kocharovskaya}}]{Coussement2002}%
  \BibitemOpen
  \bibfield  {author} {\bibinfo {author} {\bibfnamefont {R.}~\bibnamefont
  {Coussement}}, \bibinfo {author} {\bibfnamefont {Y.}~\bibnamefont
  {Rostovtsev}}, \bibinfo {author} {\bibfnamefont {J.}~\bibnamefont {Odeurs}},
  \bibinfo {author} {\bibfnamefont {G.}~\bibnamefont {Neyens}}, \bibinfo
  {author} {\bibfnamefont {H.}~\bibnamefont {Muramatsu}}, \bibinfo {author}
  {\bibfnamefont {S.}~\bibnamefont {Gheysen}}, \bibinfo {author} {\bibfnamefont
  {R.}~\bibnamefont {Callens}}, \bibinfo {author} {\bibfnamefont
  {K.}~\bibnamefont {Vyvey}}, \bibinfo {author} {\bibfnamefont
  {G.}~\bibnamefont {Kozyreff}}, \bibinfo {author} {\bibfnamefont
  {P.}~\bibnamefont {Mandel}}, \bibinfo {author} {\bibfnamefont
  {R.}~\bibnamefont {Shakhmuratov}}, \ and\ \bibinfo {author} {\bibfnamefont
  {O.}~\bibnamefont {Kocharovskaya}},\ }\href@noop {} {\bibfield  {journal}
  {\bibinfo  {journal} {Physical Review Letters}\ }\textbf {\bibinfo {volume}
  {89}},\ \bibinfo {pages} {107601} (\bibinfo {year} {2002})}\BibitemShut
  {NoStop}%
\bibitem [{\citenamefont {Gheysen}\ and\ \citenamefont
  {Odeurs}(2006)}]{Gheysen2006}%
  \BibitemOpen
  \bibfield  {author} {\bibinfo {author} {\bibfnamefont {S.}~\bibnamefont
  {Gheysen}}\ and\ \bibinfo {author} {\bibfnamefont {J.}~\bibnamefont
  {Odeurs}},\ }\href@noop {} {\bibfield  {journal} {\bibinfo  {journal}
  {Physical Review B}\ }\textbf {\bibinfo {volume} {74}},\ \bibinfo {pages}
  {155443} (\bibinfo {year} {2006})}\BibitemShut {NoStop}%
\bibitem [{\citenamefont {R\"ohlsberger}\ \emph {et~al.}(2010)\citenamefont
  {R\"ohlsberger}, \citenamefont {Schlage}, \citenamefont {Sahoo},
  \citenamefont {Couet},\ and\ \citenamefont {R\"uffer}}]{Rohlsberger2010}%
  \BibitemOpen
  \bibfield  {author} {\bibinfo {author} {\bibfnamefont {R.}~\bibnamefont
  {R\"ohlsberger}}, \bibinfo {author} {\bibfnamefont {K.}~\bibnamefont
  {Schlage}}, \bibinfo {author} {\bibfnamefont {B.}~\bibnamefont {Sahoo}},
  \bibinfo {author} {\bibfnamefont {S.}~\bibnamefont {Couet}}, \ and\ \bibinfo
  {author} {\bibfnamefont {R.}~\bibnamefont {R\"uffer}},\ }\href@noop {}
  {\bibfield  {journal} {\bibinfo  {journal} {Science}\ }\textbf {\bibinfo
  {volume} {328}},\ \bibinfo {pages} {1248} (\bibinfo {year}
  {2010})}\BibitemShut {NoStop}%
\bibitem [{\citenamefont {Raczy\'nski}\ \emph {et~al.}(2009)\citenamefont
  {Raczy\'nski}, \citenamefont {Zaremba}, \citenamefont {Zieli\'nska-Kaniasty},
  \citenamefont {Artoni},\ and\ \citenamefont {La~Rocca}}]{Raczynski2009}%
  \BibitemOpen
  \bibfield  {author} {\bibinfo {author} {\bibfnamefont {A.}~\bibnamefont
  {Raczy\'nski}}, \bibinfo {author} {\bibfnamefont {J.}~\bibnamefont
  {Zaremba}}, \bibinfo {author} {\bibfnamefont {S.}~\bibnamefont
  {Zieli\'nska-Kaniasty}}, \bibinfo {author} {\bibfnamefont {M.}~\bibnamefont
  {Artoni}}, \ and\ \bibinfo {author} {\bibfnamefont {G.~C.}\ \bibnamefont
  {La~Rocca}},\ }\href@noop {} {\bibfield  {journal} {\bibinfo  {journal}
  {Journal of Modern Optics}\ }\textbf {\bibinfo {volume} {56}},\ \bibinfo
  {pages} {2348} (\bibinfo {year} {2009})}\BibitemShut {NoStop}%
\end{thebibliography}%

\end{document}